\documentclass[10pt, conference]{IEEEtran}
\usepackage[colorlinks]{hyperref}
\usepackage{amssymb}
\usepackage{algorithm}
\usepackage{algorithmic}
\usepackage{xcolor}
\usepackage{mathrsfs}
\usepackage{blkarray}
\usepackage{amsthm}

\ifCLASSINFOpdf
  \usepackage[pdftex]{graphicx}
  \graphicspath{{./figures/}}
  \DeclareGraphicsExtensions{.pdf,.jpeg,.png}
\else
\fi

\usepackage{amsfonts}
\usepackage{amsmath}
\usepackage{caption}
\usepackage{subcaption}
\usepackage[inline]{enumitem}

\usepackage{algorithm}
\usepackage{algorithmic}
\usepackage{xcolor}
\usepackage{mathrsfs}
\usepackage{blkarray}
\usepackage{amsthm}

\newtheorem{theorem}{Theorem}

\usepackage{circledsteps}

\ifCLASSINFOpdf
\else
\fi
\usepackage{amsmath}
\usepackage{amsfonts}
\begin{document}
%
\title{Reconfigurable Intelligent Surface for OFDM Radar Interference Mitigation}
%
%
%

\author{\IEEEauthorblockN{{Ali Parchekani}, {Milad Johnny}, and {Shahrokh Valaee} }
\IEEEauthorblockA{Department of Electrical and Computer Engineering, \\
University of Toronto, Canada \\
 Email: a.parchekani@mail.utoronto.ca, milad.johnny@utoronto.ca, and valaee@ece.utoronto.ca}}

\maketitle

\begin{abstract}
This paper introduces a method to reduce interference in OFDM radar systems through the use of reconfigurable intelligent surfaces (RIS). The method involves adjusting the RIS elements to diminish interference effects and improve the clarity of the desired signal. A neural network framework is established to optimize the configurations of the RIS, aiming to lower the power from unwanted sources while enhancing the target signal. The network produces settings that focus on maximizing the signal at the intended angle. Utilizing a convolution-based approach, we illustrate the effective tuning of RIS elements for interference mitigation and the creation of nulls in the direction of interference, resulting in a better signal-to-interference-and-noise ratio (SINR). Simulations confirm the effectiveness of the proposed method in a radar context, demonstrating its capability to enhance target detection while reducing interference.
\end{abstract}


%
\IEEEpeerreviewmaketitle

\section{Introduction}
Automotive radars are essential components in modern vehicles, offering durability in adverse weather and the ability to assess both distance and velocity. OFDM radar \cite{ofdmradar} is widely used in automotive systems as it enables joint communication and sensing \cite{ofdmsensing}. However, interference among radars, especially OFDM radars, can degrade performance. Therefore, mitigating interference is crucial.

Reconfigurable Intelligent Surface (RIS)  is a promising technology that can modify signal amplitude and phase \cite{phasechange}, with versatility in diverse environments \cite{huichencommag}. RIS has been applied to Integrated Sensing and Communications (ISAC) \cite{posture}, energy-efficient communications \cite{javadkalbasi-globecom-2023}, positioning \cite{ammous-2024}, and 
radar \cite{dualradar}. It has been shown to enhance radar resolution \cite{otfspaper} and improve performance in non-line-of-sight (non-LOS) scenarios \cite{radarnonlos}, making it a viable solution for radar improvement \cite{ammouswcnc}.

Radar interference   depends on the power and duration of the interfering radar \cite{chalmers}. Various approaches have been proposed to reduce interference. For instance, \cite{compressedsensing} uses compressed sensing to mitigate interference on affected subcarriers, while \cite{compressedFMCW} applies it to reduce interference from Frequency-modulated continuous-wave (FMCW) radars. In \cite{rsofdm}, the authors propose repeated-symbol OFDM (RS-OFDM) to shape interference cooperatively, though this requires specific radar symbol generation and detection algorithms.

In this paper, we propose using RIS to mitigate interference without imposing restrictions on radar symbols. We assume interference affects all subcarriers, causing noise-like behavior. A neural network-based method is proposed to adjust RIS settings to amplify the target signal and suppress interference. Additionally, a convolution-based method combines RIS configurations to generate new interference-reducing patterns.

\section{Channel Model}

This section presents the signal model for OFDM radar with RIS. Consider a setup with a transceiver antenna on the roof of a car, and a RIS installed on the car's hood. The antenna transmits an OFDM signal consisting of $N$ subcarriers, each with $M$ symbols. The subcarrier frequencies are $f_n = f_c^{(v)} + n \Delta f$, where $f_c$ is the carrier frequency and $\Delta f$ is the subcarrier spacing. By setting $T_{\mathrm{sym}} = 1 / \Delta f$, the subcarriers remain orthogonal. To mitigate inter-symbol interference (ISI), a cyclic prefix of duration $T_{\mathrm{cp}}$ is added to each symbol, yielding a total symbol duration of $T = T_{\mathrm{sym}} + T_{\mathrm{cp}}$. The transmitted signal from the antenna is: \begin{equation} s(t) = \sum_{n,m} d^{(v)}(n,m) e^{j 2 \pi (f_c^{(v)} + n \Delta f) t} \text{rect}\left( \frac{t - mT}{T}\right). \end{equation}

The transmitted signal is reflected by a target at distance $R$, with relative velocity $v$ and angle $\theta$, resulting in a time delay $\tau = 2R/c$ and Doppler shift $\nu = 2v/c$. The signal then impinges on the RIS, with the antenna spacing of $\lambda/2$, where $\lambda = c / f_c$ is the wavelength. The received signal at the antenna, after reflection from the RIS, is: 
\begin{align} 
s_{r, \mathrm{RIS}}(t) = \sum_{l,n,m} c_{m,l} r(n,l,\theta) e^{-j 2\pi \frac{d_l}{\lambda_n}} s_{n,m}(t-\tau + t \nu), 
\end{align} 
where $r(n,l,\theta) = e^{-j2\pi \frac{\lambda}{2\lambda_n} l \cos(\theta)}$ 
with $\lambda_n = c/f_n$, $d_l$ as distance of $l$th RIS element to antenna,
and $c_{m,l}$ is the RIS configuration at time $m$. The received signal is: 
\begin{equation}
\label{target_ris}
s_{r, RIS}(t) = \sum_{n} \boldsymbol{s}_n \circ \boldsymbol{C}^T \boldsymbol{b}_n(\theta) 
\end{equation} 
where $\boldsymbol{s}_n$ represents the OFDM symbols at time $t$, $\circ$ denotes element-wise scalar multiplication, $\boldsymbol{b}_n(\theta) = [r(n,0,\theta) e^{-j 2\pi \frac{d_0}{\lambda_n}}, \ldots, r(n,L,\theta) e^{-j 2\pi \frac{d_L}{\lambda_n}}]$ , and $\boldsymbol{C}^T$ is the RIS reflection coefficient matrix, with each column representing the coefficient vector of the RIS at a specific time slot.

Next, consider an interfering OFDM radar. Assuming it operates at the same frequency, its transmitted signal is:
\begin{align}
s_I(t) &= \sum_{n,m} d^{(i)}(n,m) e^{j 2 \pi (f_c^{(v)} + n \Delta f) t} \text{rect}\left( \frac{t - mT}{T}\right). 
\end{align}
The interfering signal  is received via the RIS as: 
\begin{equation}
s_{r, I}(t) = \sum_{n,m,l} c_{m,l} e^{-j 2\pi \frac{d_i}{\lambda_n}} r(n,l,\theta) s_{I,n,m}(t - \tau_i + t \nu_i). 
\end{equation}
The received signal from the interference is expressed as: 
\begin{equation} 
s_{r,I}(t) = \sum_{n} \boldsymbol{s}_{I,n} \circ \boldsymbol{C}^T \boldsymbol{b}_n(\theta_i). \end{equation} 
The total received signal at the receiver is: 
\begin{equation}
s_{r}(t) = s_{r, RIS}(t) + s_{r,I}(t) + n(t), \end{equation} 
where $n(t)$ is the additive noise. To eliminate the direct path and constant components, we subtract signals from two consecutive frames, assuming $\boldsymbol{C}_t = - \boldsymbol{C}_{t + 1}$.

Finally, the location and velocity of the target are estimated by removing the cyclic prefix, applying a Fourier transform, and dividing the result by the transmitted symbols. The received signal at the $n$-th subcarrier of the $m$-th  symbol is: \begin{align} 
y[n,m] &= \alpha e^{-j 2\pi n \Delta f \tau} e^{j 2\pi f_c \nu m T}\nonumber\\
&+ \alpha_i \frac{d^{(i)}(n,m)}{d^{(v)}(n,m)} e^{-j 2\pi n \Delta f \tau_i} e^{j 2\pi f_c \nu_i m T} + z[n,m], 
\end{align}
where $z[n,m]$ is the noise term. The goal is to use the RIS configurations to increase the target signal power while reducing the effect of the interference. 

\section{Problem Description \& Proposed Method} \label{sec:problem}


In this section, we consider the problem of utilizing RIS to mitigate interference effects and propose a method for adjusting the RIS elements.
As discussed in the previous section, the received signal will contain a term from the interfering radar. By applying the Fourier transform and the inverse-Fourier transform, a range-velocity map (rv-map) is formed to find the peaks corresponding the target delay and Doppler shifts \cite{rvmap}. The signal of the interference radar is:
\begin{align}
    y^{(i)}[\hat{\tau}, \hat{\nu}] =& \sum_{m = 0}^{M - 1} \sum_{n = 0}^{N - 1} \alpha_i \frac{d^{(i)}(n,m)}{d^{(v)}(n,m)} 
    e^{-j 2\pi n \Delta f \tau_i}  \\ \nonumber &~~~\times e^{j 2\pi f_c \nu_i m T}  e^{j 2 \pi n \Delta f \hat{\tau}} e^{-j 2 \pi m T f_c \hat{\tau}}.
\end{align}
The above term will affect all the indices of $\hat{\tau}$ and $\hat{\nu}$, corrupting the estimation process. The effect can be considered similar to a noise behavior but with greater power since the power of the interference radar is typically higher than the target. Hence, it is important to select the configurations of the RIS such that, the received signal will have lower interference power.

\begin{figure*}
\centering
    \begin{subfigure}{.32\linewidth}
        \centering
        \includegraphics[width = 0.95 \linewidth]{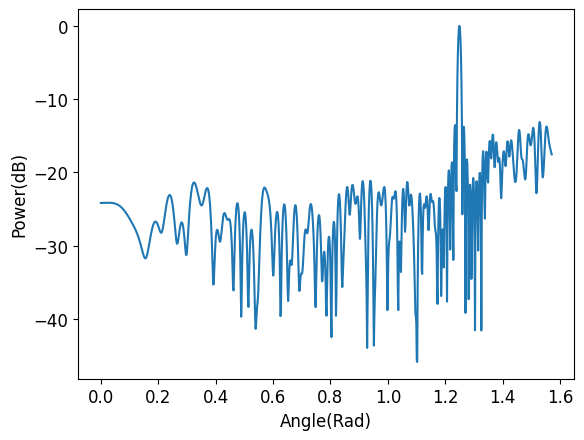}
        \caption{Normalized power pattern of $\boldsymbol{C}_p$}
        \label{fig:peak}
    \end{subfigure}
    \begin{subfigure}{.32\linewidth}
        \centering
        \includegraphics[width = 0.95 \linewidth]{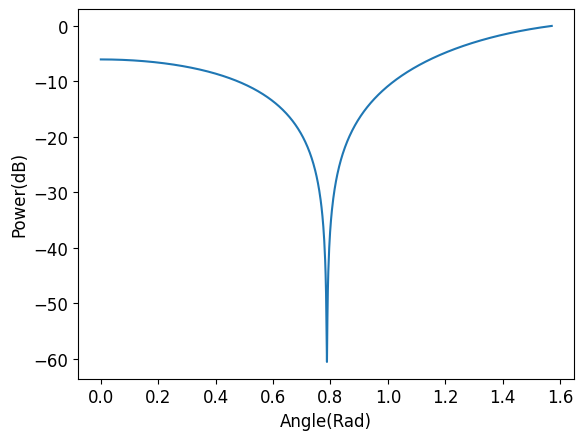}
        \caption{Normalized power pattern of $\boldsymbol{C}_n$}
        \label{fig:notch}
    \end{subfigure}
     \begin{subfigure}{.32\linewidth}
        \centering
        \includegraphics[width = 0.95 \linewidth]{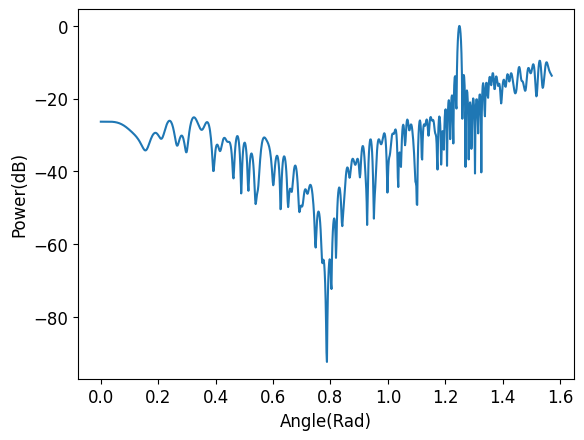}
        \caption{Combination of two patterns}
        \label{fig:combination}
    \end{subfigure}%
    \caption{Normalized power pattern of RIS configurations}
    \label{}
\end{figure*}

Let us consider the power received from a general angle $\phi$ in (\ref{target_ris}) by assuming the power of the symbols to be 1. The received power can be written as:
\begin{equation}
    P_{r, RIS}(\phi)  = \sum_{n} \lVert \boldsymbol{C}^T \boldsymbol{b}_n(\phi) \rVert_2^2.
\end{equation}
To reduce the effect of the interference and increase the power of the target, one can write the signal to interference and noise ratio (SINR). Hence, by tuning the configurations of the RIS, we can reduce the effect of the interference and increase the power of the target signal. Therefore, we have:
\begin{equation} \label{opteqn}
    \hat{\boldsymbol{C}} = \mbox{arg} \max_{\boldsymbol{C}}  \frac{\sum_{n} \lVert \boldsymbol{C}^T \boldsymbol{b}_n(\theta)\rVert_2^2}{\sum_{n}\lVert \boldsymbol{C}^T \boldsymbol{b}_n(\theta_i)\rVert_2^2 + \sigma^2}.
\end{equation}
By finding  the optimal configurations, we can reduce the effect of the interference and increase the power of the target signal.


\subsection{Proposed Method}

To find a solution for (\ref{opteqn}), we propose to use a neural network.
Consider a neural network that can be trained with an input angle $\theta_t$ and provide an output of configurations, $\boldsymbol{C}_p$, for a RIS with $L_t$ elements such that there is a peak at the given angle. 
To achieve this, a simple network with multiple linear layers and non-linear activation function such as the hyperbolic tangent can be considered. This network can be trained with the loss function $L = \frac{1}{\lVert \boldsymbol{C}_p^T \boldsymbol{b}_n (\theta_t) \rVert_2^2}$.
The goal is to find configurations for RIS that reduce the effect of the interference at any given angle $\theta_n$. Consider a RIS with $L_n$ elements, by exploring the formula $\boldsymbol{C}_n^T \boldsymbol{b}_n(\theta_n)$ one can create configurations that put the signal $\boldsymbol{b}_n(\theta_n)$ in the null space of the $\boldsymbol{C}_n$. 
By using the above network, 
one can create two configurations, $\boldsymbol{C}_p$ creating a peak at the desired angle $\theta_t$ and $\boldsymbol{C}_n$ forming a notch at the angle of interference $\theta_n$. The goal is to combine these two configurations such that a peak at $\theta_t$ and a notch at $\theta_n$ is formed.

To combine these two configurations, let us consider the received signal terms. We can assume that the distance of the RIS elements and the receiver antenna is negligible as RIS and the receiver antenna can be considered co-located. Alternatively, It is possible to consider the phase shifts of the subcarriers equal to each other as the subcarrier spacing is negligible compared to the carrier frequency and embed the phase shifts into the configurations of the RIS. The terms can be considered over the various RIS elements as:
\begin{equation}
    s_{r, \mathrm{RIS}}(t) = \sum_{l,n,m}   c_{m,l} e^{-j2\pi \frac{\lambda}{\lambda_n}l \cos(\theta)} s_{n,m}(t-\tau + t \nu).
\end{equation}
Let $\omega = 2\pi \frac{\lambda}{\lambda_n} \cos(\theta)$, for a specific carrier $n$ and angle $\theta$, and at a given sample $m$, the pattern of RIS can be written as:
\begin{equation}
    s_{r, \mathrm{RIS}}(t) = \sum_{l}  c_{m,l} e^{-j\omega l}
\end{equation}

\begin{figure}
    \centering
    \includegraphics[width=0.6\linewidth]{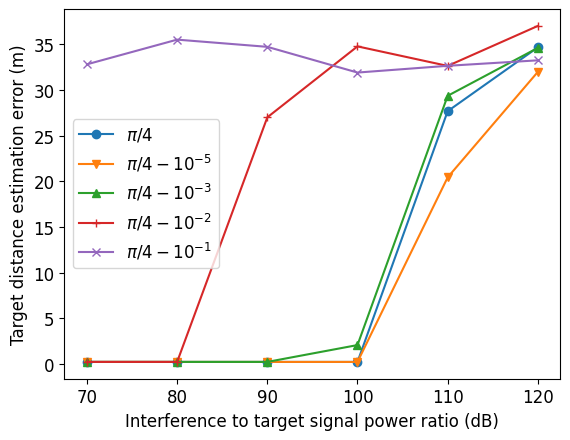}
    \caption{Distance estimation error based on interference power and target power ratio.}
    \label{fig:error_1}
\end{figure}

\begin{theorem} \label{thm}
If ${\bf c}_1 = \left[{c_{11}, c_{12}, \dots, c_{1L_1}}\right]$ and ${\bf c}_2 = \left[{c_{21}, c_{22}, \dots, c_{2L_2}}\right]$ are two RIS configurations with patterns $c_1(\theta)$ and $c_2(\theta)$, respectively, then the convolution ${\bf c}_1 * {\bf c}_2$ has the pattern $c_1(\theta)c_2(\theta)$. 
\end{theorem}

\begin{figure*}
\centering
    \begin{subfigure}{0.3\linewidth}
        \centering
        \includegraphics[width = 0.95 \linewidth]{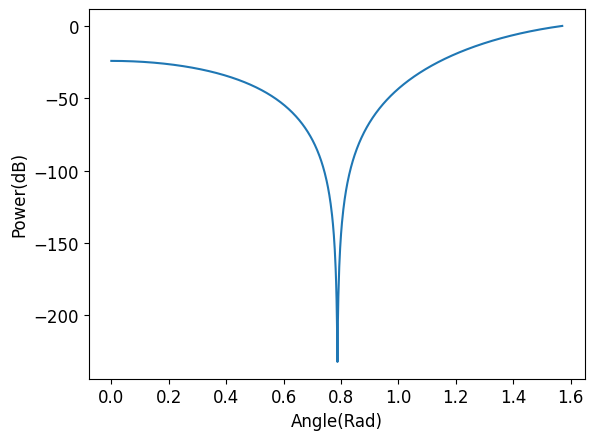}
        \caption{Power pattern for $\epsilon = 0$.}
        \label{fig:notch_4_0}
    \end{subfigure}
    \begin{subfigure}{.3\linewidth}
        \centering
        \includegraphics[width = 0.9 \linewidth]{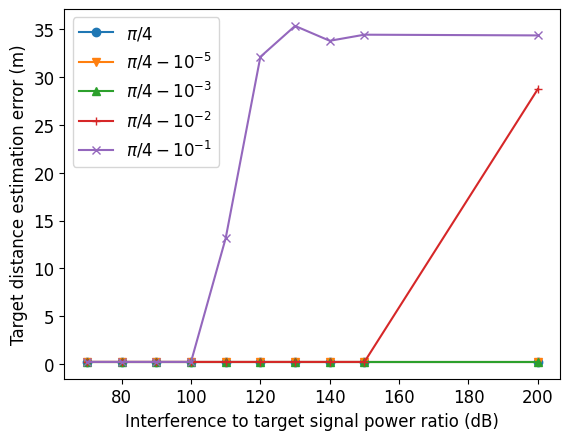}
        \caption{Estimation error for $\epsilon = 0$.}
        \label{fig:error_4_0}
    \end{subfigure}
    \begin{subfigure}{.3\linewidth}
        \centering
        \includegraphics[width = 0.95 \linewidth]{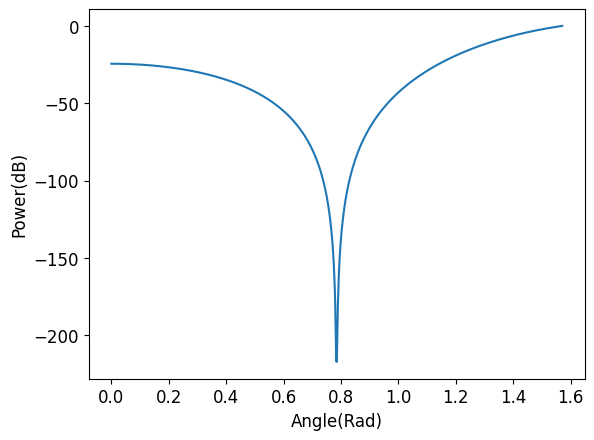}
        \caption{Power pattern for $\epsilon = 1E-3$}
        \label{fig:notch_4_1e-3}
    \end{subfigure}
    \begin{subfigure}{.3\linewidth}
        \centering
        \includegraphics[width = 0.91 \linewidth]{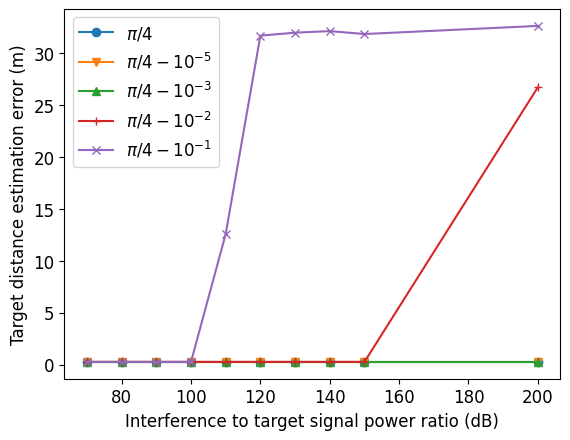}
        \caption{Estimation error for $\epsilon = 1E-3$.}
        \label{fig:error_4_1e-3}
    \end{subfigure}
     \begin{subfigure}{.3\linewidth}
        \centering
        \includegraphics[width = 0.95 \linewidth]{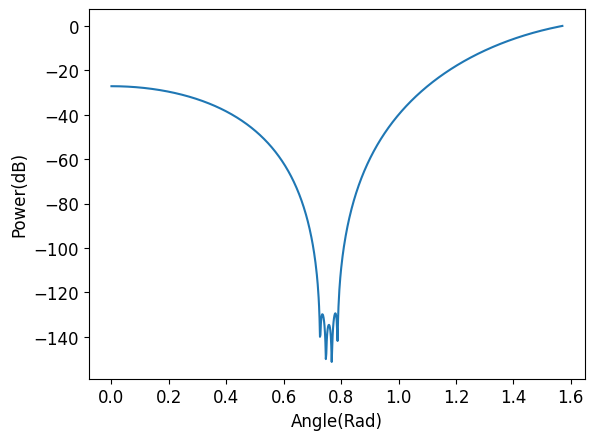}
        \caption{Power pattern for $\epsilon = 1E-2$}
        \label{fig:notch_4_1e-2}
    \end{subfigure}
    \begin{subfigure}{.3\linewidth}
        \centering
        \includegraphics[width = 0.91 \linewidth]{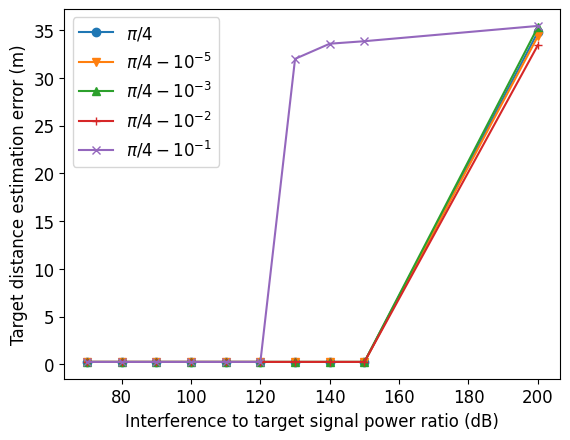}
        \caption{Estimation Error for $\epsilon = 1E-2$.}
        \label{fig:error_4_1e-2}
    \end{subfigure}%
    \caption{Normalized power pattern of multiple notches and the distance estimation error.}
    \label{fig:all_patterns}
\end{figure*}

\begin{IEEEproof}
Without loss of generality, assume that ${\bf c}_1$ and ${\bf c}_2$ are two arrays of infinite length, where the additional coefficients beyond their original lengths are zeros. We aim to prove that the pattern of the convolution ${\bf c}_1 * {\bf c}_2$ is the product of the individual patterns $c_1(\theta)$ and $c_2(\theta)$. 
The pattern of the convolution RIS can be written as:
\begin{equation}
\sum_{q=-\infty}^{\infty} \left( \sum_{p=-\infty}^{\infty} c_{1p} c_{2(q - p)} \right) e^{-j\omega q}.
\end{equation}
Let us change the order of summation and substitute $k = q - p$:
\small
\begin{align}
&\sum_{p=-\infty}^{\infty} c_{1p} \left( \sum_{q=-\infty}^{\infty}  c_{2k}  e^{-j\omega (k + p)}\right) \\ 
&= \sum_{p=-\infty}^{\infty} c_{1p} e^{-j\frac{2\pi d}{\lambda} p \cos(\theta)} \left(\sum_{k=-\infty}^{\infty} c_{2k} e^{-j\omega k}\right) \\
&= \sum_{p=-\infty}^{\infty} c_{1p} e^{-j\omega p} c_2(\theta) = c_2(\theta) c_1(\theta).
\end{align}
\normalsize
Thus, we have shown that the pattern of ${\bf c}_1 * {\bf c}_2$ is the product $c_1(\theta)c_2(\theta)$, completing the proof.
\end{IEEEproof}

Using Theorem~\ref{thm}, one can multiply the corresponding patterns of $\boldsymbol{C}_p$ and $\boldsymbol{C}_n$ to keep the peak and notch locations. The corresponding configurations for $L = L_p + L_n - 1$ is:
\begin{equation} 
\label{eq:convolve}
    \boldsymbol{C} = \boldsymbol{C}_p * \boldsymbol{C}_n,
\end{equation}
where the convolution is applied over each time frame of the configurations.


\section{Simulations} \label{sec:simul}

In this section, the numerical simulation results are presented. Let a radar system operate at $f_c = 77$ GHz with the available bandwidth of $B = 200 \mbox{MHz}$. The number of subcarriers and their corresponding number of samples are $N = 100$ and $M = 50$. The modulation symbols are randomly chosen from QPSK. RIS is located at the coordinate center and the tranceiver radar antenna is located at $(-0.5m, 0.5m)$. Consider a target at the angle of $\theta = \frac{2 \pi}{5}$ and an interference radar operating with the same characteristics located at the angle of $\theta_i =\frac{\pi}{4}$. Let us assume that $L_p = 200$ elements are used to train the neural network with 6 linear layers and activation function hyperbolic tangent function to design the peak at the angle of $\theta$ resulting in the configuration $\boldsymbol{C}_p$. $L_n = 2$ elements are used to create a notch at the angle of the interference radar, $\theta_i$, with the configuration $\boldsymbol{C}_n = [1, e^{-j \pi \cos(\theta_i)}]$. Fig.~\ref{fig:peak} and Fig.~\ref{fig:notch} show the beampattern over various angles formed by the RIS elements. Fig.~\ref{fig:combination} shows the combination of the RIS patterns using (\ref{eq:convolve}). It can be seen that the resulting pattern combines the individual patterns and creates a peak and a notch at the desired angles.


To analyze the error in target range estimation, we employ the maximum likelihood estimation (MLE) technique as outlined in \cite{rvmap}. We evaluate the error under various interference-to-target power ratios and report its impact on the accuracy of target range estimation. Additionally, we examine the sensitivity of the antenna's notch with respect to the angle of interference. Fig.~\ref{fig:error_1} presents the range estimation error over a range of angles around $\theta_i$. As depicted, the estimation error increases significantly with higher interference power levels. This error primarily stems from the imperfect alignment of the interference signal with the antenna's null, causing residual interference to act as a noise term due to the symbol mismatch between the interference and target radars. Moreover, after a certain threshold, the error exhibits fluctuations. This behavior is attributed to the substantial power leakage from the interference, which introduces randomness into the estimation process and leads to variability in the observed error.

To reduce the error, it is possible to increase the number of elements involved in creating the notch. For that, we can combine several notches around $\theta_i$ and with spacing of $\epsilon$. Fig.~\ref{fig:all_patterns} shows the notch pattern as well as the distance estimation errors for combining 4 notches using 5 elements of RIS. It can be seen that increasing the number of elements further suppresses the interference power. Hence, resulting in lower estimation error even at high power ratios. Increasing the spacing of the notches, $\epsilon$, reduces the suppression but increases the robustness against uncertainty in estimation of the $\theta_i$ as it can suppress a wider range of angles but with lower power. Therefore, depending on the level of certainty, there exists a trade-off between the power suppression and the range of the suppressed angles.

\section{Conclusion} \label{sec:conclusion}
This paper presented a method for optimizing RIS configurations to reduce interference in OFDM radar systems. By applying neural networks to generate RIS element configurations, we successfully created peaks at target angles and notches at interference angles. Our approach, leveraging convolution to combine RIS patterns, demonstrated significant improvements in SINR and target detection accuracy. The proposed technique is efficient, scalable, and adaptable to various environmental conditions, offering a robust solution for radar applications where interference reduction is critical.


%



\ifCLASSOPTIONcaptionsoff
  \newpage
\fi



%



\bibliographystyle{IEEEtran}
\bibliography{bibtex/bib/IEEEexample}

\begin{thebibliography}{10}
\providecommand{\url}[1]{#1}
\csname url@samestyle\endcsname
\providecommand{\newblock}{\relax}
\providecommand{\bibinfo}[2]{#2}
\providecommand{\BIBentrySTDinterwordspacing}{\spaceskip=0pt\relax}
\providecommand{\BIBentryALTinterwordstretchfactor}{4}
\providecommand{\BIBentryALTinterwordspacing}{\spaceskip=\fontdimen2\font plus
\BIBentryALTinterwordstretchfactor\fontdimen3\font minus \fontdimen4\font\relax}
\providecommand{\BIBforeignlanguage}[2]{{%
\expandafter\ifx\csname l@#1\endcsname\relax
\typeout{** WARNING: IEEEtran.bst: No hyphenation pattern has been}%
\typeout{** loaded for the language `#1'. Using the pattern for}%
\typeout{** the default language instead.}%
\else
\language=\csname l@#1\endcsname
\fi
#2}}
\providecommand{\BIBdecl}{\relax}
\BIBdecl

\bibitem{ofdmradar}
Y.~L. Sit, C.~Sturm, L.~Reichardt, T.~Zwick, and W.~Wiesbeck, ``The {OFDM} joint radar-communication system: An overview,'' 2011.

\bibitem{ofdmsensing}
S.~D. Liyanaarachchi, C.~B. Barneto, T.~Riihonen, and M.~Valkama, ``Joint {OFDM} waveform design for communications and sensing convergence,'' in \emph{ICC 2020 - 2020 IEEE International Conference on Communications (ICC)}, 2020, pp. 1--6.

\bibitem{phasechange}
Z.~Wu, Y.~Ra'di, and A.~Grbic, ``Tunable metasurfaces: A polarization rotator design,'' \emph{Phys. Rev. X}, vol.~9, p. 011036, Feb 2019.

\bibitem{huichencommag}
H.~Chen, H.~Kim, M.~Ammous, G.~Seco-Granados, G.~C. Alexandropoulos, S.~Valaee, and H.~Wymeersch, ``{RISs} and sidelink communications in smart cities: The key to seamless localization and sensing,'' \emph{IEEE Communications Magazine}, vol.~61, no.~8, pp. 140--146, 2023.

\bibitem{posture}
J.~Hu, H.~Zhang, B.~Di, L.~Li, K.~Bian, L.~Song, Y.~Li, Z.~Han, and H.~V. Poor, ``Reconfigurable intelligent surface based {RF} sensing: Design, optimization, and implementation,'' \emph{IEEE Journal on Selected Areas in Communications}, vol.~38, no.~11, pp. 2700--2716, 2020.

\bibitem{javadkalbasi-globecom-2023}
M.~Javad-Kalbasi, M.~S. Al-Abiad, and S.~Valaee, ``Energy efficient communications in {RIS}-assisted {UAV} networks based on genetic algorithm,'' in \emph{Proc. {IEEE} Global Communications Conference, (GLOBECOM)}, 2023.

\bibitem{ammous-2024}
M.~Ammous, H.~Chen, H.~Wymeersch, and S.~Valaee, ``Zero access points 3d cooperative positioning via {RIS} and sidelink communications,'' in \emph{arXiv:2305.08287}, 2024.

\bibitem{dualradar}
Y.~He, Y.~Cai, H.~Mao, and G.~Yu, ``{RIS}-assisted communication radar coexistence: Joint beamforming design and analysis,'' 2022.

\bibitem{otfspaper}
A.~Parchekani and S.~Valaee, ``Sensing via orthogonal time frequency space signalling and reconfigurable intelligent surface,'' in \emph{2022 IEEE 33rd Annual International Symposium on Personal, Indoor and Mobile Radio Communications (PIMRC)}, 2022, pp. 1--6.

\bibitem{radarnonlos}
A.~Aubry, A.~De~Maio, and M.~Rosamilia, ``Reconfigurable intelligent surfaces for n-los radar surveillance,'' \emph{IEEE Transactions on Vehicular Technology}, vol.~70, no.~10, pp. 10\,735--10\,749, 2021.

\bibitem{ammouswcnc}
M.~Ammous and S.~Valaee, ``{RIS}-enabled cooperative sidelink positioning under partial blockage,'' in \emph{IEEE WCNC}, 2024.

\bibitem{chalmers}
G.~K. Carvajal, M.~F. Keskin, C.~Aydogdu, O.~Eriksson, H.~Herbertsson, H.~Hellsten, E.~Nilsson, M.~Rydström, K.~Vänas, and H.~Wymeersch, ``Comparison of automotive {FMCW} and {OFDM} radar under interference,'' in \emph{2020 IEEE Radar Conference (RadarConf20)}, 2020, pp. 1--6.

\bibitem{compressedsensing}
B.~Nuss, L.~Sit, and T.~Zwick, ``A novel technique for interference mitigation in {OFDM} radar using compressed sensing,'' in \emph{2017 IEEE MTT-S International Conference on Microwaves for Intelligent Mobility (ICMIM)}, 2017, pp. 143--146.

\bibitem{compressedFMCW}
C.~Knill, B.~Schweizer, and C.~Waldschmidt, ``Interference-robust processing of {OFDM} radar signals using compressed sensing,'' \emph{IEEE Sensors Letters}, vol.~4, no.~4, pp. 1--4, 2020.

\bibitem{rsofdm}
B.~Schweizer, C.~Knill, D.~Werbunat, S.~Stephany, and C.~Waldschmidt, ``Mutual interference of automotive {OFDM} radars—analysis and countermeasures,'' \emph{IEEE Journal of Microwaves}, vol.~1, no.~4, pp. 950--961, 2021.

\bibitem{rvmap}
M.~Braun, C.~Sturm, and F.~K. Jondral, ``Maximum likelihood speed and distance estimation for {OFDM} radar,'' in \emph{2010 IEEE Radar Conference}, 2010, pp. 256--261.

\end{thebibliography}

%








\end{document}